# Autonomous Sampling and SHAP Interpretation of Deposition-Rates in Bipolar HiPIMS

Alexander Wieczorek[1,+], Nathan Rodkey[1,+,*], Jan Sommerhäuser[1], Jason Hattrick-Simpers[2,3], Sebastian Siol[1,*]

[1]Laboratory for Surface Science and Coating Technologies, Empa–Swiss Federal Laboratories for Materials Science and Technology, Ueberlandstrasse 129, Duebendorf CH-8600, Switzerland.

[2]Department of Materials Science and Engineering, University of Toronto, M5S 3E4, Toronto, ON, Canada.

[3]Acceleration Consortium, University of Toronto, M7A 2S4, Toronto, ON, Canada.

+ These authors contributed equally

* Corresponding authors: nathan.rodkey@empa.ch, sebastian.siol@empa.ch

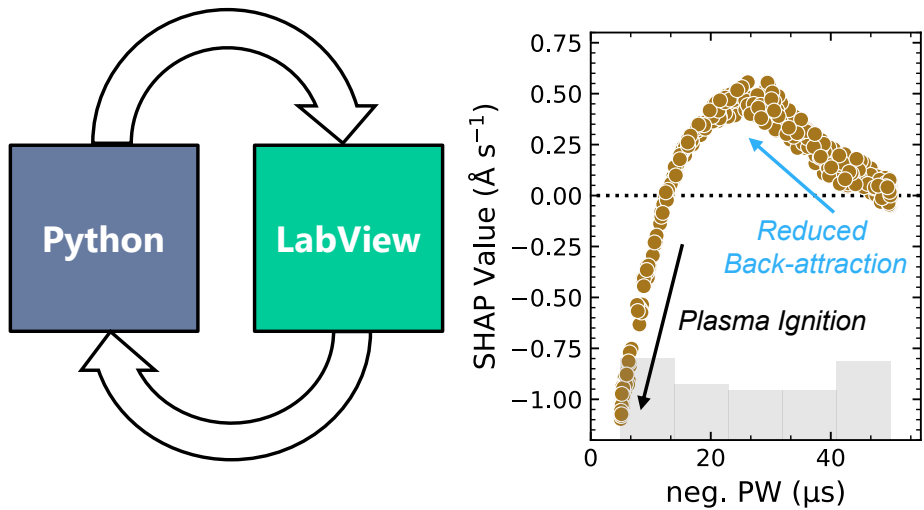

**Abstract:** High-power impulse magnetron sputtering (HiPIMS) offers considerable control over ion energy and flux, making it invaluable for tailoring the microstructure and properties of advanced functional coatings. However, compared to conventional sputtering techniques, HiPIMS suffers from reduced deposition rates. Many groups have begun to evaluate complex pulsing schemes to improve upon this, leveraging multi-pulse schemes (e.g. pre-ionization or bipolar pulses). Unfortunately, the increased complexity of these pulsing schemes has led to high-dimensionality parameter spaces that are prohibitive to classic design of experiments. In this work we evaluate bipolar HiPIMS pulses for improving deposition rates of Al and Ti sputter targets. Over 3000 process conditions were collected via autonomous Bayesian sampling over a 6-dimensional parameter space. These process conditions were then interpreted using Shapley Additive Explanations (SHAP), to deconvolute complex process influences on deposition rates. This allows us to link observed variations in deposition rate to physical mechanisms such as back-attraction and plasma ignition. Insights gained from this approach were then used to target specific processes where the positive pulse components were expected to have the highest impact on deposition rates. However, in practice, only minimal improvements in deposition rate were achieved. In most cases, the positive pulse appears to be detrimental when placed immediately after the neg. pulse which we hypothesize relates to quenching of the afterglow plasma. The proposed workflow combining autonomous experimentation and interpretable machine learning is broadly applicable to the discovery and optimization of complex plasma processes, paving the way for physics-informed, data-driven advancements in coating technologies.

# INTRODUCTION

High-power impulse magnetron sputtering (HiPIMS) is a deposition technique where short, µs-scale voltage pulses are applied to a sputter target at low duty cycles. Consequently, high peak-current densities ($J_{pk}$) and higher plasma-densities can be achieved, resulting in a higher ionized flux fraction of the sputtered species.[1] This leads to some unique advantages, as increased ionization allows for an accelerating substrate bias to be applied, which can be used to tailor microstructure, stress, and even phase formation.[2–7] Moreover, the generally higher kinetic energy of species can be leveraged to improve adatom mobility, leading to dense, highly oriented structures, as shown recently for AlN and AlScN thin films.[8–10] However, HiPIMS processes suffer from low deposition-rates when compared to their DC or pulsed-DC analogues.

Several factors contribute to this low deposition rate, one of the most important being ion back-attraction; a phenomenon by which positively charged ions are attracted to the negatively charged sputter target, reducing their probabilities for escape.[11] Additionally, the low duty cycles common in HiPIMS processes mean that significant energy goes into ionizing the process gas at the beginning of each pulse.[1,5,7] As a result, improving deposition-rates has been of particular interest to the community, studied through a variety of techniques and methods. Popular among these is the introduction of different pulsing schemes, such as the use of mid-frequency pulses to pre-ionize the plasma,[12,13] pulse-packeting,[14–16] or bipolar signals.[17] This growing complexity mirrors a broader trend in materials processing. Unfortunately, the amount of data available in the field for these techniques is still limited and increasing pulse complexity makes classical design of experiments prohibitive.

Specifically, in the case of bipolar HiPIMS pulses, recent reports have shown a notable increase in the deposition-rate of sputtered ions. This was attributed to the idea of a positive sheath around the sputter target that would reflect ions and reduce back-attraction.[18,19] However, these results were contradicted by studies that failed to see any experimental increase in deposition rate,[20–22] as well as several studies of plasma dynamics that showed that the initial idea of an ion-reflecting sheath was not seen experimentally.[17,20,21,23] This was related to similar contradictory reporting in bipolar HiPIMS of ion acceleration onto insulating substrates which relied on the idea of this ion-reflecting sheath. Previously, *Tiron et al.*[24] demonstrated that ion acceleration towards a floating substrate is possible using bipolar pulses with proper balancing of the magnetrons; in most other cases the positive component of the pulse increases the bulk plasma potential, dropping only as it approaches the substrate sheath. [17,20–23,25,26] In such cases, no notable ion acceleration is observed as the surface potential closely matches that of the plasma.

Such contradictory reporting can partly be explained by the complexity of the process when configuring pulse shapes. A bipolar HiPIMS pulse adds three parameters to the dimensionality of the process (i.e. pos. delay, pos. pulse-width, and pos. voltage). Coupled with typical process adjustments to frequency, pulse-width, and peak-current density ($J_{pk}$) of the negative pulse, the number of permutations in the experimental design make process development exceedingly difficult through a classic design of experiments. In response to this, most reports on bipolar HiPIMS fix two of these three features, changing only the voltage or delay in their experimental design.[18,19,26] As such, to unlock the full potential of bipolar HiPIMS, there is a need for larger datasets, coupled with an efficient way to collect and interpret these datasets.

Growing demand for data is not unique to the HiPIMS community. Data rich experimentation and machine learning (ML) approaches promise to solve this timely challenge, aiming to accelerate understanding and control of processes with increasing complexity. In other fields, rapid adoption of experimental automation and ML techniques has already materialized this potential.[27–35] However, the development of self-driving or autonomous physical vapor deposition (PVD) systems has generally lagged behind, with few examples of fully autonomous PVD labs.[32] This can be attributed to the complexity involved in experimental automation of vacuum systems as well as the need for *in-vacuo* materials characterization. A more practical approach to accelerated PVD process development is to use *in-situ* process diagnostic which allows for non-intrusive, continuous data collection[36–38] and/or batch processing.[39,40] Generally, these experimental setups rely on Bayesian optimization to efficiently sample process windows, however, high-dimensionality processes (e.g. bipolar HiPIMS) require additional, complex interpretation tools.

Fortunately, developments in ML interpretation published in 2017 by *Lundberg et al*. showed that Shapley Additive Explanations (SHAP), a statistical formulation originating from game theory, could be used to interpret feature influence in complex machine-learning models.[41] SHAP explanations work by calculating the expected contribution of each feature (e.g. process parameter) to an output (e.g. deposition rate), shown as a deviation from the model's base value (i.e. the mean prediction). This is also referred to as the marginal contribution of the given feature and has the same physical units as the model output, enabling quantitative interpretation of feature effects. SHAP has become a widely adopted method for model interpretability and has been increasingly applied to process design in materials science.[42–49]

In this work, we use Bayesian statistics to efficiently sample the deposition-rate in a bipolar HiPIMS discharge and use SHAP to interpret the resulting datasets. The resulting dataset represents the deposition-rates of >3000 process conditions for bipolar HiPIMS pulses. We choose Al and Ti metals for the study to compare materials with relatively lower (e.g. Al) or higher (e.g. Ti) ionized flux fraction. In addition, Ti typically exhibits higher amounts of doubly, or triply charged ions (e.g. Ti), which we expect to have stronger interactions with the positive pulse components.[50] SHAP-interpretation of the data demonstrated that generally, the positive components of a bipolar HiPIMS pulse have no meaningful impact on the deposition-rate, as proven for both Al and Ti datasets. Subsequently, we targeted specific processes in a high $J_{pk}$ dataset where the positive pulse components were expected to have the largest impact, however, potential gains in deposition rate were minimal. Instead, many processes showed reduced deposition rates with the introduction of a pos. pulse, which we hypothesize relates to quenching of the afterglow plasma. Finally, the large amount of data represented in this work allow for a powerful visual of back-attraction and plasma ignition effects in HiPIMS as they relate to deposition rates.

**EXPERIMENTAL METHODS**

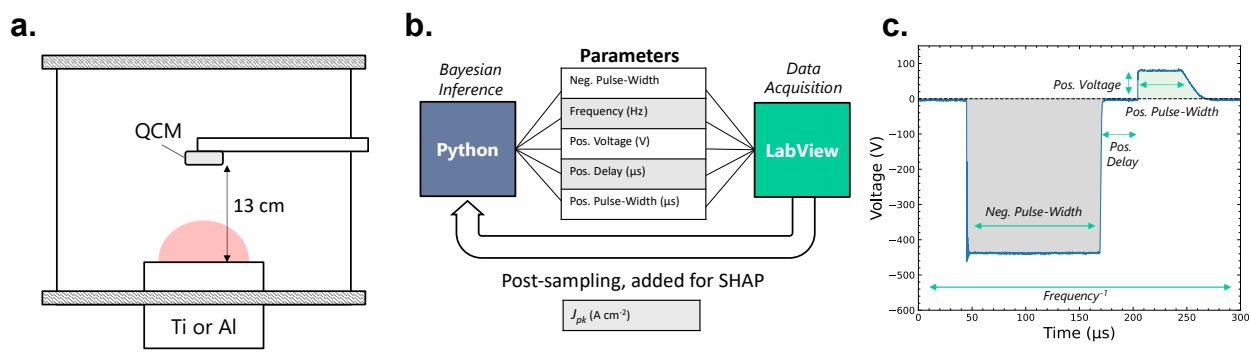

**Figure 1**. Schematic of **a.** the experimental setup, involving a QCM placed at a distance of 13 cm from a Ti or Al sputter target, **b.** the feedback loop used between a Python-based Bayesian algorithm, which communicates with LabView to run different process parameters, and **c.** an example of a bipolar HiPIMS pulse.

Experiments were carried out in a custom-built sputter chamber using an unbalanced 3" magnetron (A330, *AJA International Inc.*) in a coplanar, sputter-up geometry. The sputter targets were sourced from *Kurt J. Lesker* at a purity >99.99% for both Al and Ti targets. All depositions were performed with 50 sccm Ar flow, regulated to 0.5 Pa working pressure by throttling a gate-valve. A quartz-crystal monitor (QCM) was used to measure deposition rates, placed at a distance of 13 cm from the sputter target (shown schematically in **Figure 1a**). Datasets were collected using constant power on the sputter target, requiring multiple datasets to be collected at different power densities (2.63, 4.4, and 5.48 W cm$^{-2}$) to access high peak-current density ($J_{pk}$) over a wide pulse-width range of 5 – 300 µs. The boundary conditions and accessible $J_{pk}$ range for each of these datasets is summarized in **Table S1**. In short, a low and high negative pulse-width (neg. PW) was collected for both Al and Ti, constrained to frequencies between 500 – 5000 Hz for the low neg. PW datasets (5 – 100 µs), and frequencies of 200 - 800 Hz for high neg. PW datasets (100 – 300 µs). Additionally, a low duty-cycle dataset was collected for both Al and Ti to access higher $J_{pk}$ (1.5 – 1.8 Acm$^{-2}$), constrained to a PW of 5 – 50 µs and a duty cycle (frequency * PW) range of 1.2 – 3.75%. Additionally, to prevent overlapping between the positive and negative pulse components, the lower and upper bounds of the *pos. Delay* and *PW* components were chosen accordingly, set to a range from 0 – 40 µs.

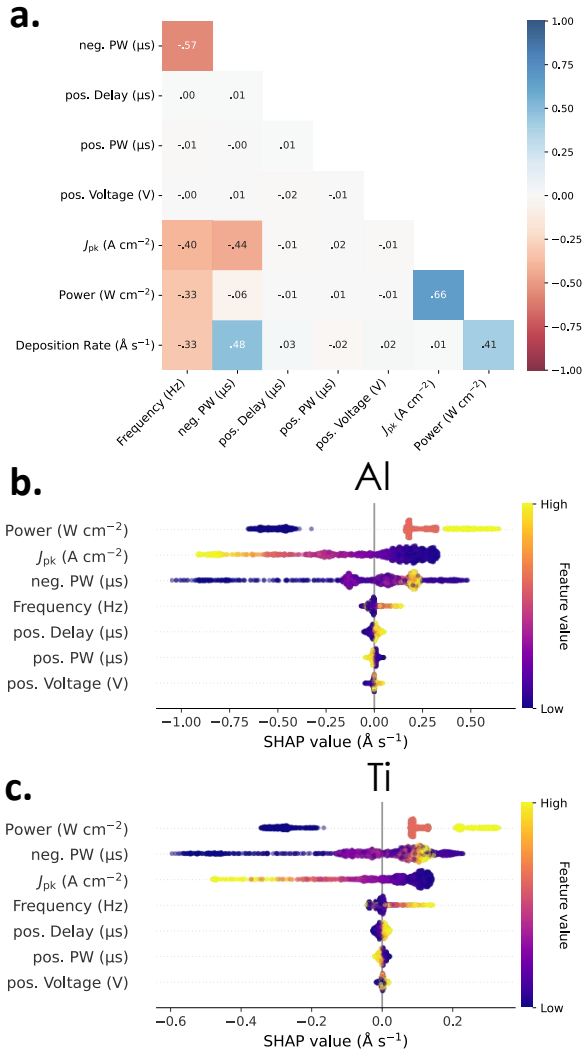

**Figure 2.** Global overview of the bipolar HiPIMS data sets. **a.** Spearman correlation matrix of all process parameters as well as the resulting deposition rate. For **b.** Al, and **c.** Ti, a SHAP beeswarm plot with the mean of the datasets marked as a solid line, and the SHAP value on the x-axis indicating the predicted deviation from the model mean caused by each process parameter. Process parameters are ordered by feature importance.

The sputter chamber, workflow, and an example of a bipolar HiPIMS pulse pattern are shown schematically in **Figure 1**. The sputter chamber is controlled via software implemented in LabView 2024 Q1, which interfaces with Python 3.12.11 for the adaptive sampling. This sampling uses Bayesian statistics implemented through the open-source package BayBE 0.13.1 which is based on BoTorch 0.14.0.[51] A Gaussian Proccess Regressor (GPR) based on a Matérn kernel was used as the surrogate model. As default in BayBE, hyperparameters were optimized using the L-BFGS-B algorithm via the fit_gpytorch_mll routine from BoTorch.[52] The Bayesian algorithm was run in exploration mode by using the posterior standard deviation to minimize uncertainty within the predefined bounds of the parameter-space. Once the process conditions are set by the Python routine, LabView executes it, first stabilizing the process for 10 s, followed by measurement of the deposition-rate, done by integrating the total deposition until 35.1 and 21.7 Å were deposited for Al and Ti respectively (total mass deposited on the QCM was the same for both Al and Ti). The corresponding process parameters are then logged, including the oscilloscope waveforms of each process, averaged over 10 scans. Peak-currents are then determined from the waveforms using a *Python* algorithm. All *Python* code used in this work is included in the *Supporting Information*.

Collected data was then interpreted using a game theory model, Shapley Additive Explanations (SHAP).[41,53] All SHAP interpretations shown in this work are done using an Exact Explainer unless otherwise stated; interfaced via an open-source package, *SHAP* version 0.50.0.[53] Notably, $J_{pk}$ was not controlled in the Bayesian exploration (only measured), and the uncertainty was thus not minimized across that space. The uncertainty across the parameter space was evaluated, including $J_{pk}$, by calculating the mean standard deviation over a uniformly spaced mesh of ~390,000 points, shown in **Figure S1** for all 6 datasets. The mean standard deviation does not reduce significantly past 100 measurements.

**RESULTS AND DISCUSSION**

### Global Overview of Data

We start our investigation with a global analysis of all datasets recorded in this study. Notably, this includes studies on both Al and Ti metal targets. Analysis of the Spearman correlation coefficients, as shown in **Figure 2a**, suggests low pairwise correlation between most process parameters. Additionally, individual spearman correlation matrices for Al and Ti only datasets are shown in **Figure S2**. Of note, *Power Density* showed a high correlation value of 0.73 with $J_{pk}$ but was left in the analysis to allow for a generalized overview of the six datasets, which were performed at different power densities. Following this, SHAP was used to estimate feature importance.

This global overview is shown in **Figure 2b** and **c** as beeswarm plots, summarizing the three datasets of Al in **Figure 2b** and of Ti in **Figure 2c**. In a beeswarm plot, features are ranked by importance according to the absolute sum of their SHAP values, with the model mean set to zero, shown as a black reference line. Each point represents a unique process, for which the SHAP values are the attributed deviation from the model mean caused by that parameter. The sum of all SHAP values in a process sum to 0 and SHAP values retain the units of the model output (Å s$^{-1}$).

Among all parameters, power density emerges as the most influential factor driving deposition rate. However, the broad spread of SHAP values indicate overlapping effects and limited separability between their contributions; an expected outcome given the sparse sampling of power density values. Following the power density, unipolar HiPIMS process parameters ($J_{pk}$, *neg. PW*, and *Frequency*) rank next in feature importance. Trends observed in the unipolar components are consistent with established observations. High values of $J_{pk}$, drop the deposition rate, attributed to the correlation of high $J_{pk}$ to higher ion counts and a corresponding reduction in deposition rates from the increased influence of back-attraction to these ions. Next, the *neg. PW* trends are more nuanced, suggesting some local maximum, with the highest contributions to deposition-rate coming from lower *neg. PW*. This is also an expected reaction relationship relating to back-attraction as low *neg. PW* values help mitigate back-attraction effects, as shown by *Shimizu et al.*[11] This relationship is discussed in greater detail in the following section. Finally, the *Frequency* has the smallest effect amongst the unipolar pulse parameters, showing an expected increase in deposition rate at higher *Frequency* values. This is also a well-reported phenomenon, where the higher background plasma densities at higher frequencies allow for more efficient plasma ignition and greater sputter yield.[12,13,54] Finally, parameters associated with the pos. pulse components of bipolar HiPIMS display low absolute SHAP values, suggesting only a minor contribution to the overall deposition rate. Individual beeswarm plots for all six datasets are provided in **Figure S3**.

### Interactions and Correlations

To gain deeper insight into the underlying structure of the data, scatter plots of SHAP values are used to examine correlations, interaction effects, and general trends throughout this work. In **Figure 3a**, a spread in the frequency is observed, depending on the neg. PW. In SHAP, this is called feature interaction; the features' importance depends on the value of another. More specifically, as the neg. PW increases, the impact of frequency decreases. This is seen as well in **Figure 3b**, which gives an idea of the spread of data over the parameter space. Here, the neg. PW is plotted against Frequency and its SHAP value on the color-scale.

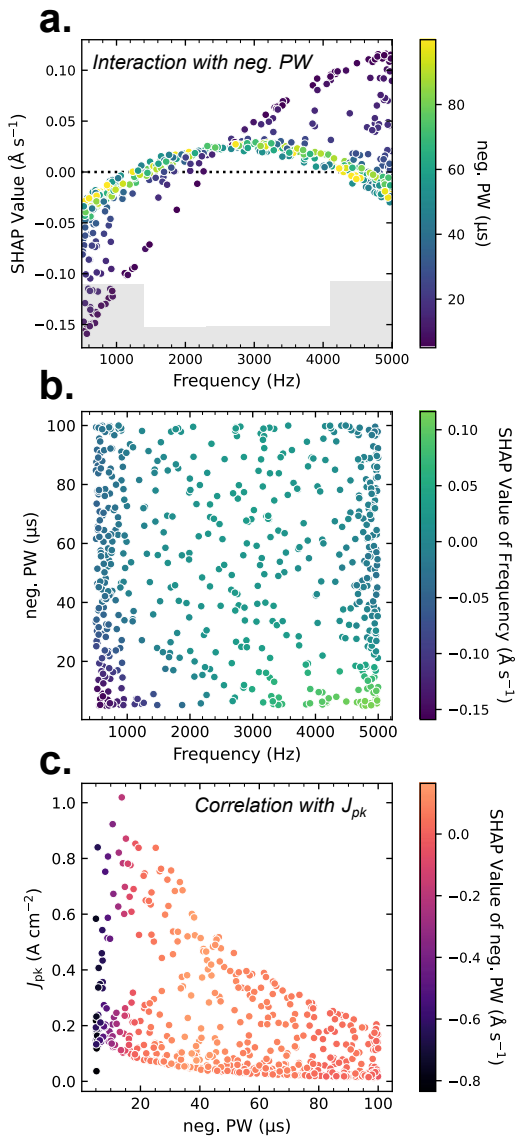

**Figure 3.** SHAP analysis for a low pulse-width, Al dataset is shown above. In **a.** dependence plot, highlighting an interaction effect between Frequency and neg. PW. The model mean of the dataset is marked as a dotted line. The distribution of data collected over the parameter space is visualized by a grey bar chart. The distribution of sampled points is then shown for **b.** neg. PW vs. Frequency and **c.** $J_{pk}$ vs. neg. PW. The color scale shows the SHAP value of the relevant feature, shown on the x-axis. These highlights correlations found in the dataset as $J_{pk}$ tends to increase at lower neg. PW, a result of power-control used during the sputter deposition.

In **Figure 3c** a correlation is observed, with higher $J_{pk}$ at lower neg. PW values. This comes from the use of power control during data sampling, which limits the voltage applied to the sputter target (and thus $J_{pk}$). This was needed to keep $J_{pk}$ below values where arcing or plasma instabilities are observed. The correlation of $J_{pk}$ to neg. PW is mirrored in the frequency as well (shown in **Fig. S4**). Such correlations make it difficult to ascertain where interactions observed in **Figure 3a,b** come from, as they can be a result of spurious attribution. In this case, the true interaction with frequency may be with $J_{pk}$, but is mediated through the neg. PW because of underlying correlations in the dataset.

Correlations can sometimes be handled by conditional SHAP analysis techniques, such as permutation or partition explainers, which instead of applying the empirical mean to "missing" values when creating coalitions during SHAP analysis (e.g. exact explainer) will randomly permute missing features (permutation explainer) or sample expected values based on the created coalitions (partition explainer). We have found in the study of these datasets that conditional explanations, shown in **Figure S5**, do not separate the interactions seen here between frequency and neg. PW. This may indicate some validity to those interactions as frequency and neg. PW together describe the off-time in the HiPIMS discharge, which is important for background plasma conditions relevant to the deposition rate.

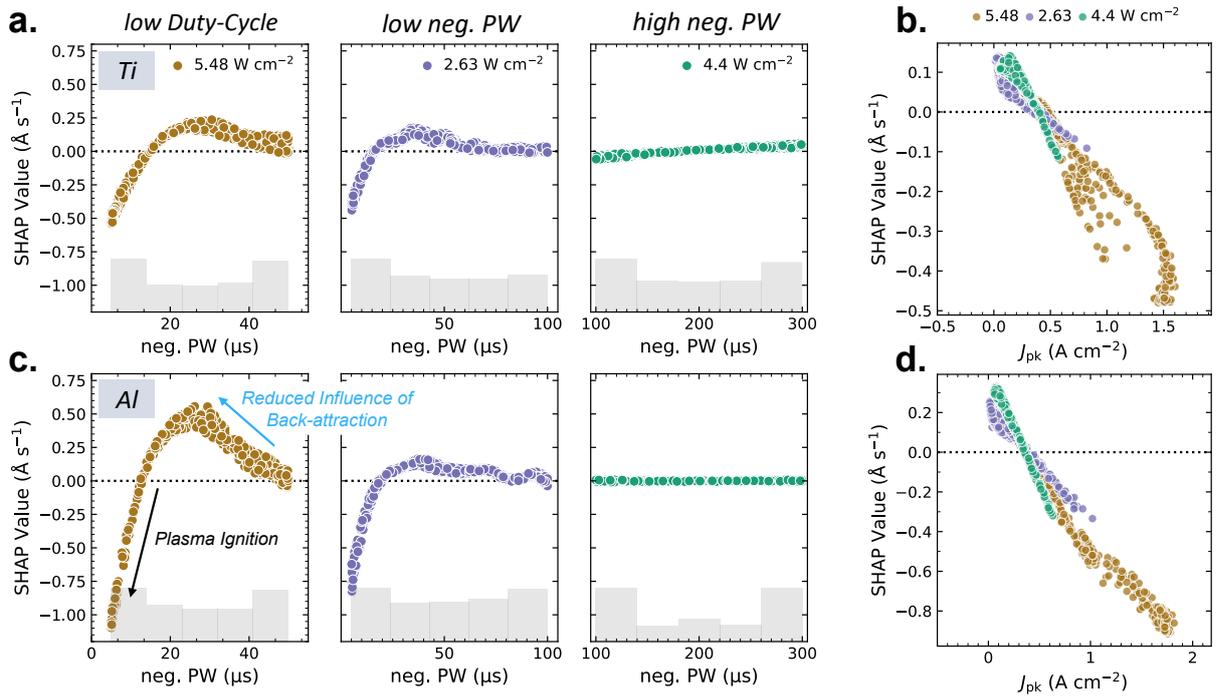

**Figure 4.** SHAP interpretations of the neg. PW (**a,c**) and $J_{pk}$ (**b,d**) are shown for all six datasets collected in this work. This includes datasets for both Ti and Al collected at varying power densities (indicated by color) and neg. PW ranges; datasets are labeled on the top as low duty-cycle, low neg. PW, and high neg. PW for which additional information on their boundary conditions is included in **Table S1**. The model mean of each dataset is marked by a dotted line and the SHAP value on the y-axis indicates the predicted deviation from this mean. Three datasets of each **a.** Ti and **c.** Al are shown, where the SHAP interpretation of the neg. PW shows a notable increase in impact, attributed to the reduced influence of back-attraction. This is seen prominently in the low duty-cycle datasets. This prominence is attributed to the higher mean $J_{pk}$ seen in the low duty-cycle dataset. The distribution of datapoints is visualized by the grey bar chart at the bottom of each graph. Additionally, a sharp decrease is observed at low neg. PW, attributed to energy loss during plasma ignition. As $J_{pk}$ is an important feature for determining deposition-rates, we show for each **b.** Al and **c.** Ti the combined three datasets. SHAP-interpretation on this global dataset shows a near-linear trend with deposition rate.

## Visualization of Back-Attraction and Plasma-Ignition

Detailed SHAP analysis of parameters linked to high feature importance give insights into physics informed trends. As a reminder, and as described in the *Experimental Methods* section, datasets were collected at discrete power densities to access a reasonable range of $J_{pk}$ over a wide neg. PW range. The details of the boundary conditions and range accessible in each dataset are summarized in **Table S1**.

In **Figure 4**, we can see the SHAP scatter plots for all datasets involved in this study, covering >3000 process conditions. In **Figure 4a** for Ti and **4c** for Al, the influence of the neg. PW becomes prominent for datasets with a low neg. PW range (5 - 100 µs), with a visible peak around 25 - 30 µs for both Al and Ti datasets. This sudden increase in importance of the neg. PW is a known phenomenon in HiPIMS; it relates to the attraction of sputtered positive ions to the negatively charged sputter target, referred to as back-attraction. This causes a severe reduction in deposition-rates and is reported in several theoretical papers,[17,20,21,23] as well as experimentally verified by *Shimizu et al.*[11] When a voltage pulse ends, back-attraction ceases as the target potential returns to 0, and positively charged ions experience enhanced deposition-rates. This is seen as an increase in deposition-rate as pulses become shorter as the fraction of deposition happening after the voltage pulse ends becomes more significant. The increase in prominence of this effect for the low duty-cycle datasets is attributed to the higher $J_{pk}$ and resulting higher ion counts, strongly affected by back-attraction. The mean $J_{pk}$ of datasets is shown in **Table S2** for reference. For the high neg. PW datasets, this is not observed as the reduced back-attraction pertaining to the end of the pulse becomes insignificant to the whole.

This peak is followed by a gradual, and subsequently sharp decrease at lower neg. PW. Indications of this relationship were seen experimentally by several groups[11,55] and likely relate to the onset of plasma ignition, which can consume a significant amount of energy. This is interpreted by SHAP analysis as a negative correlation with the neg. PW, as energy spent on plasma ignition becomes proportionally more significant.

Additionally, spreading of the data noted in the low duty-cycle datasets was caused by the interaction of the neg. PW with the $J_{pk}$ (see **Figure S6**). As $J_{pk}$ is an indicator of ion concentration, and the impact of back-attraction depends on this ion-concentration, it follows that these two features interact.

Visible as well in **Figure 4a** and **c** for the low neg. PW datasets are minor oscillations within the 50 – 100 µs range. Although currently unexplained, the consistency of these features across many process points indicates a physical origin rather than statistical noise. As such, the complex plasma dynamics potentially underpinning this feature warrant further investigations.

Finally, for concise presentation of the data we choose to include a global overview of $J_{pk}$ trends in **Figure 4b** and **d** for Ti and Al datasets respectively, although individual SHAP analysis of $J_{pk}$ in these datasets can be found in **Figure S7**. This generalized overview of $J_{pk}$ highlights the effectiveness of SHAP analysis at deconvoluting dominant process parameters such as power density or neg. PW. Deposition rates follow near-linear trends with $J_{pk}$ for each individual power density.

## Local Verification of Trends Indicated by SHAP

Finally, to target high $J_{pk}$ regimes, a dataset was collected covering a range of ~0.2 – 2 A cm$^{-2}$. This was done by placing lower and upper bounds on the duty cycle of the process as opposed to the frequency, as described in the *Experimental Methods* section. These targeted datasets for Al and Ti were collected, covering a range of *neg. PW* from 5 – 50 µs, with Ti shown in **Figure 5** and Al in **Figure S3**, expecting the significance of pos. bipolar pulse components to increase in a higher $J_{pk}$ dataset (and thus greater ionization of the process); and indeed, an increase of significance was observed, seen by the beeswarm plot in **Figure 5a** or the side-by-side comparisons in **Figure S3**. Note that changes in the importance of SHAP parameters are expected, as the exact explainers used in this work rely on empirical means when creating coalitions. . As HiPIMS process parameters tend to interact heavily with $J_{pk}$ (and thus ionization flux fraction), we have included in **Table S2** the mean $J_{pk}$ of each dataset, as a reference when visualizing these datasets.

In **Figure 5a**, the beeswarm suggests a significant impact of the pos. pulse components. Namely, to improve deposition-rates over the model mean, the pos. Delay and Voltage should be increased, and the pos. PW decreased. This suggests that the pos. pulse components should be pushed far away from the HiPIMS pulse, and for the shortest duration possible, seen across all 6 datasets collected in this work (**Fig. S3**) and contradicts the conventional thinking that the pos. pulse should be placed immediately after the neg. pulse.

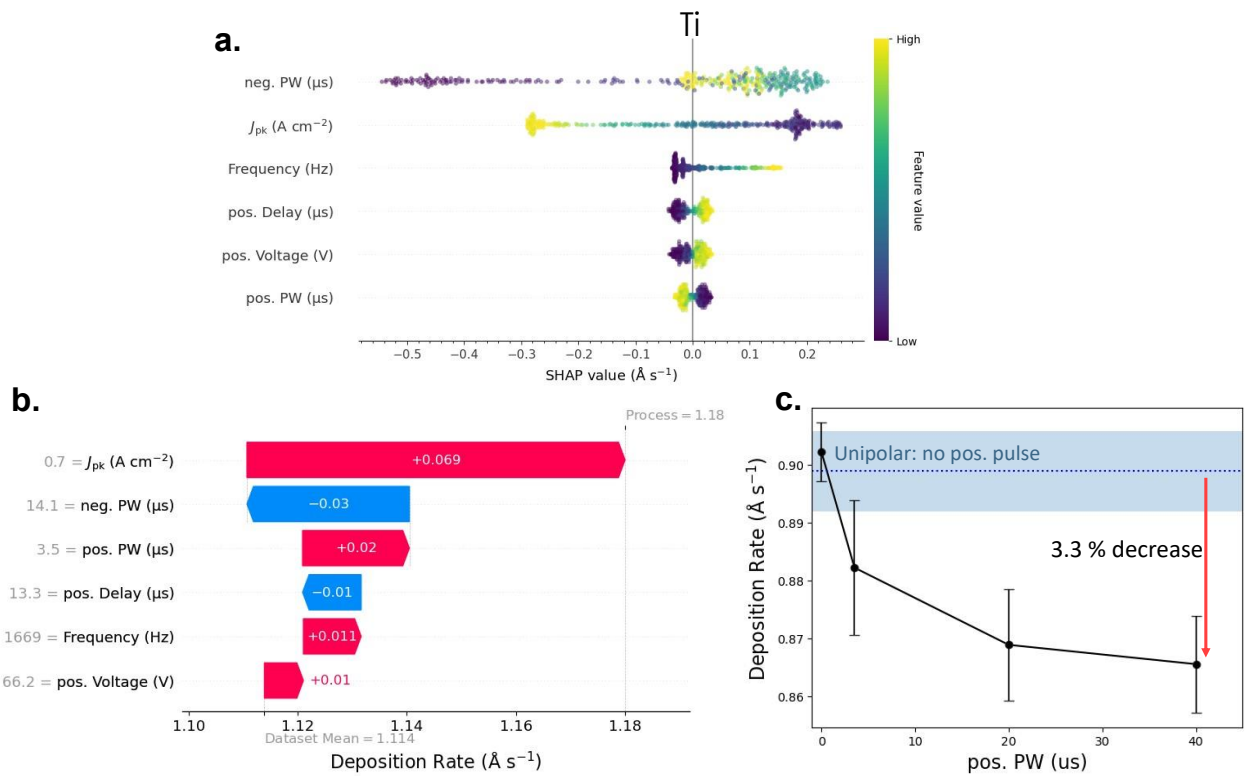

**Figure 5.** Above for Ti, and in **Figure S3** for Al, a dataset acquired under low duty-cycle (high $J_{pk}$) conditions. Pos. pulse components become visibly more significant compared to the global analysis, as seen in the beeswarm plot (panel **a**). By comparing the SHAP values of the pos. pulse components with the absolute sum of SHAP values in individual processes, we identified processes where the pos. component is expected to have the highest impact. An example is shown in panel **b.** as a waterfall plot of a process heavily impacted by pos. PW. Subsequently, measurements were manually performed while fixing all process parameters except the predicted most impactful pos. pulse parameter. An example for pos. PW is shown in panel **c.** with the measurements for other pos. pulse parameters shown in **Figure S8**. The dotted blue line indicates the measured deposition rate when all pos. pulse components were turned off, with the error bars and shaded blue area representing one standard deviation shaded. Overall, the trends in the pos. pulse components measured here align with those observed in the SHAP analysis, although deviate from the estimated linearity, likely due to the sparse sampling at the edge of the parameter space.

We theorize that the significance observed here (for the pos. Delay and PW) relates to interactions with the background plasma conditions where significant deposition-rates have been observed for several hundred μs between HiPIMS pulses (i.e. during the afterglow), seen experimentally by *Mitschi et al.*.[55] Additionally, many groups report "quenching" of the plasma upon the application of a pos. pulse, noticed by onset delays in the rise of $J_{pk}$[21] as well as a drop in electron density of several orders of magnitude (over several 100 µs) after applying a positive pulse, effectively halting ionization and at times extinguishing the afterglow.[56] These complex, and heavily process-dependent plasma dynamics may explain some of the contradictory reporting seen with bipolar HiPIMS deposition-rates. The dependence of background plasmas on bipolar pulses, magnetron balancing, and many other process parameters is a wide and complex field of study.[20,57–59]

Visualized in **Figure S8**, we also note prominent interaction effects observed for each pos. pulse component (e.g. vertical spreading of the data) which arise from interactions with multiple different features and are quantified using a grid-based variance approach, summarized in **Table S3**.

By summing up the absolute SHAP values for an individual process, we can also rank which ones are most likely to be influenced by pos. pulse components. A local explanation of a process expected to be heavily impacted by pos. PW is shown as a waterfall plot in **Figure 5b**. The different process parameters are displayed on the left, in order of importance, showing their expected contributions in increasing (red) or decreasing (blue) the measurements deviation from the model mean. In **Figure 5b**, we can see that the *pos. PW* is estimated to increase the dep-rate by 0.02 Å s$^{-1}$ from the model mean and is a significant component.

Several processes exhibiting a high predicted influence of pos. pulse components were selected for manual verification in this way. The local explanations of each of these processes are included in **Figure S9**. Manual verification involved using the same parameters of the original process but sweeping the pos. pulse component of interest and measuring the deposition rates. Additionally, the process was run with all pos. pulse components turned off to get an idea of improvements over a pure unipolar case. For example, the local explanation shown in **Figure 5b** shares the same process conditions as the manually measured deposition rates shown in **Figure 5c**. In **Figure 5c**, the pos. PW was swept from 0 – 40 µs noting a continual decrease in deposition rates, followed; previously attributed to quenching of the plasma afterglow. Each point was measured 3 times, with the error bars and the shaded blue area around the unipolar case representing one standard deviation. The decrease in deposition rate when compared to a unipolar process with the same process conditions (dotted blue line) is estimated to be 3.3%. Similar measurements can be found for pos. Delay and Voltage next to their local explanations in **Figure S9**. Some improvements in deposition-rate are seen in **Figure S9** but were minimal and often within the error of the measurement.

In general, the trends observed in the beeswarm plot of **Figure 5a** are conserved, with a need for high pos. Delay, low pos. PW, and high pos. Voltage to increase dep-rates. Overall, the analysis in **Figure 5** may indicate that improvements in deposition rate are possible, but minimal using a positive HiPIMS pulse. In most cases, the positive pulse appears to be detrimental to the deposition-rate when placed immediately after the neg. pulse. We hypothesize that this behavior arises from a combination of factors: a minor, and likely short-lived, ion-reflecting sheath near the target, which enhances the deposition rate of nearby ions; premature quenching of the afterglow which reduces the residual deposition; and lower background plasma densities as a result of plasma quenching, which increases energy losses during the ionization of process gases at the beginning of each pulse.

## CONCLUSION

By effectively and autonomously sampling a large parameter space of >3000 bipolar HiPIMS process conditions, we demonstrate a data-driven approach to understanding deposition-rate improvements in bipolar HiPIMS. Using SHAP analysis, we were able to produce powerful visualizations of physical HiPIMS mechanisms such as back-attraction and plasma ignition.

Minimal improvements in deposition rate were possible, but in most cases, the positive pulse appears to be detrimental when placed immediately after the neg. pulse. We hypothesize that this behavior arises from a combination of factors the most important likely being associated with a quenching of the afterglow which reduces the residual deposition. The lower background plasma densities also increase energy losses during the ionization of process gases at the beginning of each HiPIMS pulse. Further investigation is warranted to investigate these phenomena in detail.

Overall, the results show that autonomous sampling using in-situ plasma and process diagnostics is a powerful tool to gain insight into modern plasma-based deposition processes, in highly complex parameter spaces. Comprehensive data sets facilitate process modelling in a field where results are often influenced by external factors which are hard to control and monitor, such as chamber geometry, magnetic field distribution or even precursor purity. The proposed workflow combining autonomous experimentation and interpretable ML is broadly applicable and can be expanded with relevant *in-situ* materials characterization, paving the way for physically informed, data-driven advancements in coating technologies.

## ACKNOWLEDGEMENTS

A.W. and S.S. acknowledge funding from the Strategic Focus Area–Advanced Manufacturing (SFA–AM) through the project Advancing manufacturability of hybrid organic–inorganic semiconductors for large area optoelectronics (AMYS). JHS acknowledges funding provided to the University of Toronto's Acceleration Consortium by the Canada First Research Excellence Fund (CFREF-2022-00042). A.W., J.S. and S.S. further acknowledge financial support by the ETH ORD program for the development of data management tools. S.S. and N.R. acknowledge the Swiss National Science Foundation under Projects 226588 and 10004403. Ulrich Müller, Sasa Vranjkovic and Ismail Casgar are gratefully acknowledged for their help in the technical construction and manufacturing of the deposition chamber. The authors further acknowledge Manuel Kober-Czerny, Fabio Lopes and Felix Thelen for helpful discussions.

**AUTHOR CONTRIBUTIONS**

CRediT: A.W.: Conceptualization, Data curation, Formal Analysis, Investigation, Methodology, Software, Visualization, Writing – original draft, Writing – review & editing; N.R.: Conceptualization, Data curation, Formal Analysis, Investigation, Methodology, Software, Visualization, Writing – original draft, Writing – review & editing; J.S.: Investigation, Writing – review & editing; J.H-S.: Methodology, Writing – review & editing; S.S.: Conceptualization, Funding acquisition, Methodology, Resources, Supervision, Writing – review & editing

Magnetron Sputtering. *Surf Coat Technol* 2019, *359* (11), 97–107. https://doi.org/10.1016/j.surfcoat.2018.12.079.

(27) *FULL-MAP - Acceleration Consortium*. https://www.full-map.eu/.

(28) Ikenson, B. Full Automation of Research Systems for Materials Science May Be on the Horizon. *Scilight* 2022, *2022* (2). https://doi.org/10.1063/10.0009348.

(29) Tom, G.; Schmid, S. P.; Baird, S. G.; Cao, Y.; Darvish, K.; Hao, H.; Lo, S.; Pablo-García, S.; Rajaonson, E. M.; Skreta, M.; Yoshikawa, N.; Corapi, S.; Akkoc, G. D.; Strieth-Kalthoff, F.; Seifrid, M.; Aspuru-Guzik, A. Self-Driving Laboratories for Chemistry and Materials Science. *Chem Rev* 2024, *124* (16), 9633–9732. https://doi.org/10.1021/acs.chemrev.4c00055.

(30) Chen, A.; Zhang, X.; Zhou, Z. Machine Learning: Accelerating Materials Development for Energy Storage and Conversion. *InfoMat* 2020, *2* (3), 553–576. https://doi.org/10.1002/inf2.12094.

(31) *Acceleratin Consortium - University of Toronto*. https://acceleration.utoronto.ca/.

(32) Aiba, A.; Nishio, K.; Hitosugi, T. Development of a Digital Laboratory Integrating Modular Measurement Instruments. 2025, 1–5. https://doi.org/https://doi.org/10.1039/d4dd00326h.

(33) Ishizuki, N.; Shimizu, R.; Hitosugi, T. Autonomous Experimental Systems in Materials Science. *Science and Technology of Advanced Materials: Methods* 2023, *3* (1). https://doi.org/10.1080/27660400.2023.2197519.

(34) MacLeod, B. P.; Parlane, F. G. L.; Rupnow, C. C.; Dettelbach, K. E.; Elliott, M. S.; Morrissey, T. D.; Haley, T. H.; Proskurin, O.; Rooney, M. B.; Taherimakhsousi, N.; Dvorak, D. J.; Chiu, H. N.; Waizenegger, C. E. B.; Ocean, K.; Mokhtari, M.; Berlinguette, C. P. Advancing the Pareto Front for Thin-Film Materials Using a Self-Driving Laboratory. 2021, 1–38. https://doi.org/https://doi.org/10.1038/s41467-022-28580-6.

(35) Abolhasani, M.; Kumacheva, E. The Rise of Self-Driving Labs in Chemical and Materials Sciences. *Nature Synthesis* 2023, *2* (6), 483–492. https://doi.org/10.1038/s44160-022-00231-0.

(36) Schaefer, S.; Febba, D.; Egbo, K.; Teeter, G.; Zakutayev, A.; Tellekamp, B. Rapid Screening of Molecular Beam Epitaxy Conditions for Monoclinic ($In_xGa_{1−x}$)$_2O_3$ Alloys. *J Mater Chem A Mater* 2024, *12* (9), 5508–5519. https://doi.org/10.1039/d3ta07220g.

(37) Fébba, D. M.; Talley, K. R.; Johnson, K.; Schaefer, S.; Bauers, S. R.; Mangum, J. S.; Smaha, R. W.; Zakutayev, A. Autonomous Sputter Synthesis of Thin Film Nitrides with Composition Controlled by Bayesian Optimization of Optical Plasma Emission. *APL Mater* 2023, *11* (7). https://doi.org/10.1063/5.0159406.

(38) Jarl, S.; Sjölund, J.; Frost, R. J. W.; Holst, A.; Scragg, J. J. S. Machine Learning for In-Situ Composition Mapping in a Self-Driving Magnetron Sputtering System. 2025, 1–24. https://doi.org/10.1016/j.matdes.2025.115087.

(39) Zheng, Y.; Blake, C.; Mravac, L.; Zhang, F.; Chen, Y.; Yang, S. A Machine Learning Approach Capturing Hidden Parameters in Autonomous Thin-Film Deposition. 2024, 1–8. https://doi.org/10.48550/arXiv.2411.18721.

(40) Harris, S. B.; Biswas, A.; Yun, S. J.; Roccapriore, K. M.; Rouleau, C. M.; Puretzky, A. A.; Vasudevan, R. K.; Geohegan, D. B.; Xiao, K. Autonomous Synthesis of Thin Film Materials with Pulsed Laser Deposition Enabled by In Situ Spectroscopy and Automation. *Small Methods* 2024, *8* (9), 1–11. https://doi.org/10.1002/smtd.202301763.
19

Wieczorek et al., Empa, 2026

# *Supporting Information:* Autonomous Sampling and SHAP Interpretation of Deposition-Rates in Bipolar HiPIMS

*Alexander Wieczorek[1,+], Nathan Rodkey[1,+,*], Jan Sommerhäuser[1], Jason Hattrick-Simpers[2,3], Sebastian Siol[1,*]*

[1]Laboratory for Surface Science and Coating Technologies, Empa–Swiss Federal Laboratories for Materials Science and Technology, Ueberlandstrasse 129, Duebendorf CH-8600, Switzerland.

[2]Department of Materials Science and Engineering, University of Toronto, M5S 3E4, Toronto, ON, Canada.

[3]Acceleration Consortium, University of Toronto, M7A 2S4, Toronto, ON, Canada.

*+ These authors contributed equally*

*\* Corresponding authors: nathan.rodkey@empa.ch, sebastian.siol@empa.ch*

**Table S1** – Each dataset was collected using power control, setting lower and upper bounds for some process conditions during the exploration phase. These are process conditions that were considered during the Bayesian optimization and whose uncertainty was minimized over those bounds. For *low neg. PW* and *high neg. PW* datasets, lower and upper bounds were set on the neg. PW, frequency, and pos. pulse components. For *low duty-cycle* datasets, lower and upper bounds were set on the neg. PW, duty-cycle, and pos. pulse components. In all cases $J_{pk}$ was not considered during the exploration and its error not minimized. Instead it was added for post-analysis.

| Dataset Label | Boundary Conditions for Each Parameter (lower bound, upper bound) | | | | | | | |
|---|---|---|---|---|---|---|---|---|
| | $J_{pk}$ (A cm$^{-2}$) | neg. PW (µs) | Frequency (Hz) | Duty Cycle (%) | pos. Voltage (V) | pos. Delay (µs) | pos. PW (µs) | Power Density (W cm$^{-2}$) |
| Al – low duty-cycle | (0.28 – 1.82) | (5 – 50) | (240 – 7500) | (1.2 – 3.75) | (0 – 100) | (1.5 – 40) | (0 – 40) | 5.48 |
| Al – low neg. PW | (0.01 – 1.02) | (5 – 100) | (500 – 5000) | (0.25 – 50) | (0 – 100) | (1.5 – 40) | (0 – 40) | 2.63 |
| Al – high neg. PW | (0.05 – 0.63) | (100 – 300) | (200 – 800) | (2 – 24) | (0 – 100) | (1.5 – 40) | (0 – 40) | 4.4 |
| Ti – low duty-cycle | (0.25 – 1.60) | (5 – 50) | (240 – 7500) | (1.2 – 3.75) | (0 – 100) | (1.5 – 40) | (0 – 40) | 5.48 |
| Ti – low neg. PW | (0.01 – 0.82) | (5 – 100) | (500 – 5000) | (0.25 – 50) | (0 – 100) | (1.5 – 40) | (0 – 40) | 2.63 |
| Ti – high neg. PW | (0.05 – 0.57) | (100 – 300) | (200 – 800) | (2 – 24) | (0 – 100) | (1.5 – 40) | (0 – 40) | 4.4 |

**Table S2** – Empirical mean of peak-current density for Al and Ti low duty cycle datasets.

| Datasets | Al – low neg. PW | Al – high neg. PW | Al – low duty-cycle | Ti – low neg. PW | Ti – high neg. PW | Ti – low duty-cycle |
|---|---|---|---|---|---|---|
| $J_{pk}$ (A cm$^{-2}$) | 0.175 | 0.205 | 0.995 | 0.163 | 0.205 | 0.816 |

**Table S3** – Grid-based variance is used to quantify interaction effects. Each feature is split into 6 bins and the weighted variance is evaluated across columns and rows. This is shown below for the SHAP explanation of a low duty cycle Ti dataset. The evaluated features (pos. pulse components) are shown on the left with the weighted variance across rows and columns shown with the corresponding features.

|  | **Variance Across this Feature** | | | | | |
|---|---|---|---|---|---|---|
| **Evaluated Feature** | $J_{pk}$ (A cm$^{-2}$) | neg. PW (µs) | Frequency (Hz) | pos. Voltage (V) | pos. Delay (µs) | pos. PW (µs) |
| pos. Voltage (V) | 0.036 | 0.098 | 0.033 | - | 0.012 | 0.019 |
| pos. Delay (µs) | 0.016 | 0.032 | 0.028 | 0.037 | - | 0.009 |
| pos. PW (µs) | 0.023 | 0.030 | 0.030 | 0.012 | 0.017 | - |

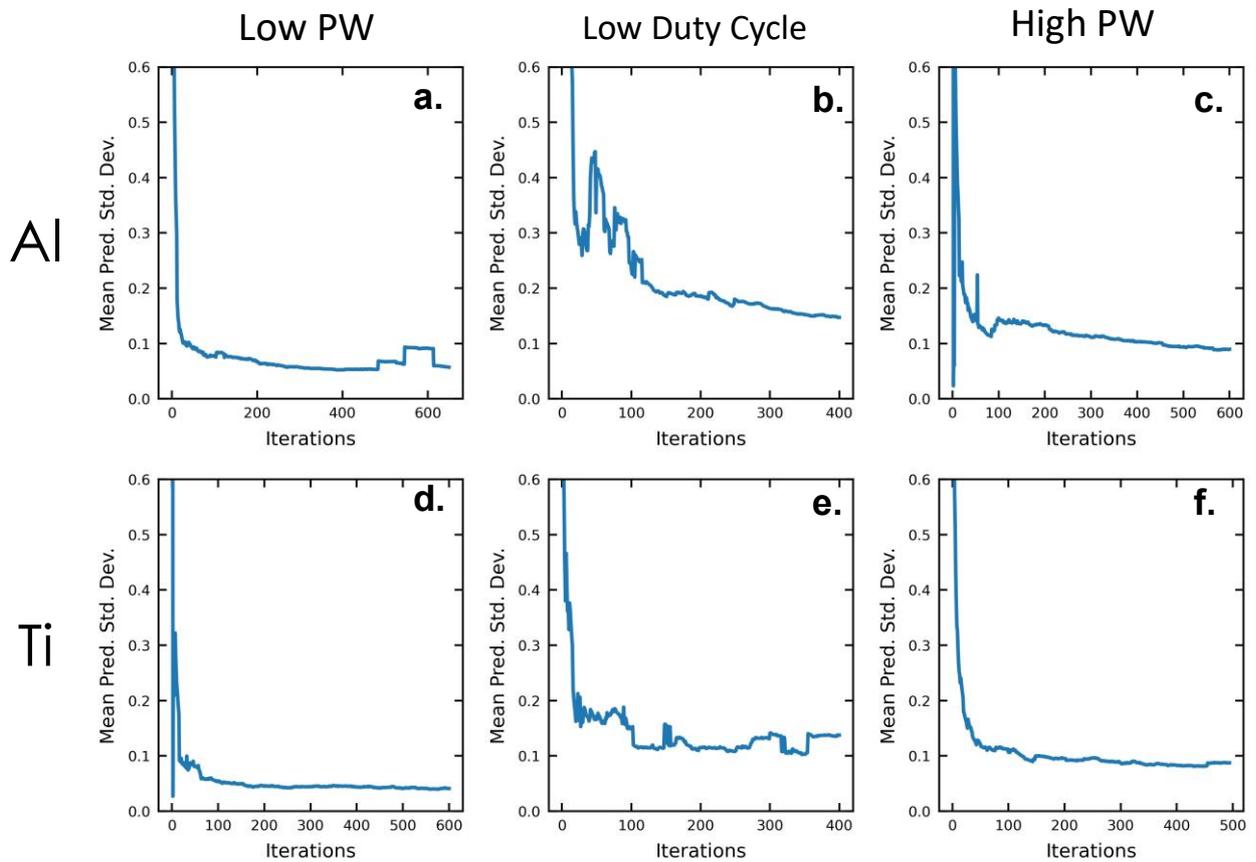

**Figure S1.** Mean standard deviations of datasets over iterations were evaluated by evenly splitting each feature into 5 bins to create testing data. Measurements of each dataset were then added manually, one at a time, evaluating the mean standard deviation across the testing data with the updated surrogate model. This is shown for Al (**a-c**) and Ti (**d-f**) datasets for the low pulse-width (**a,d**), low duty cycle (**b,e**), and high pulse-width (**c,f**) datasets, as described in the *Experimental Methods*. In general, the mean standard deviation is minimized after ~100 – 200 iterations, indicating high-quality sampling of the parameter space.

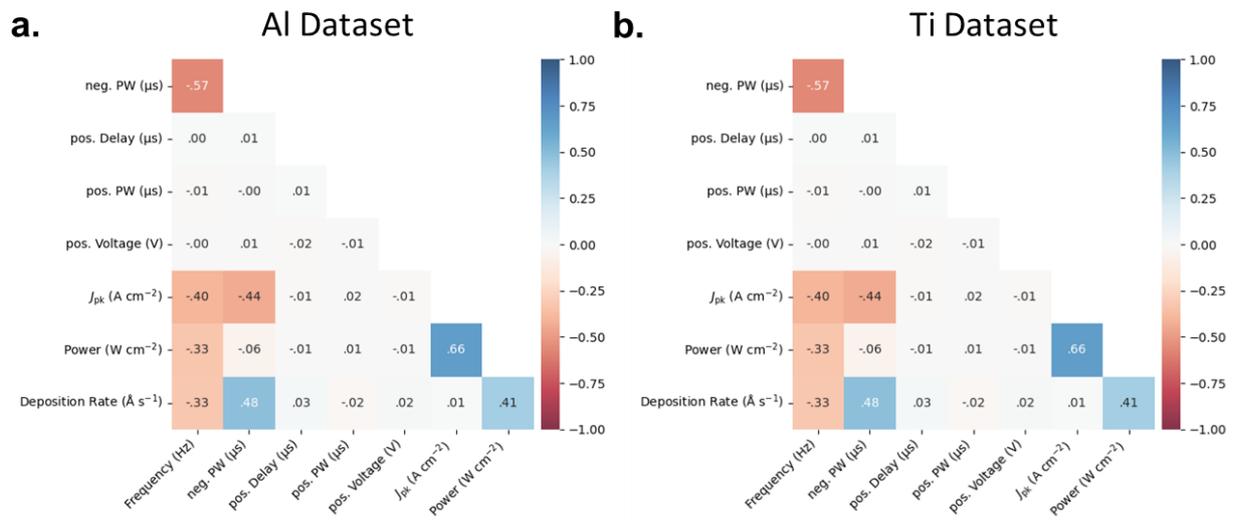

**Figure S2.** Spearman correlation matrices separated by metal-type for **a.** Al and **b.** Ti.

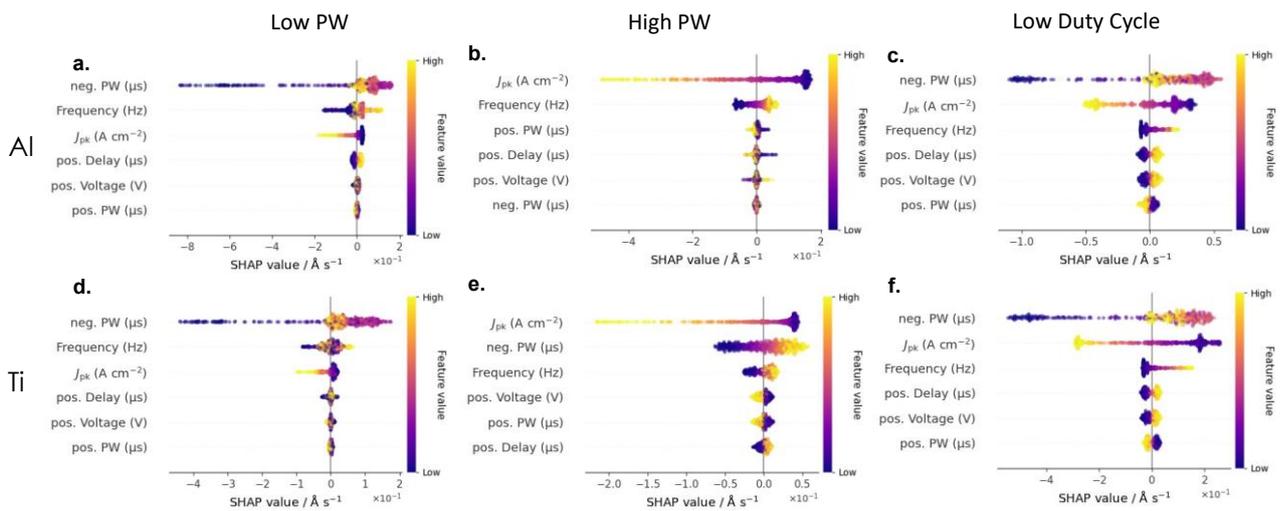

**Figure S3.** Beeswarm plots for all Al (**a-c**) and Ti (**d-f**) datasets are provided.

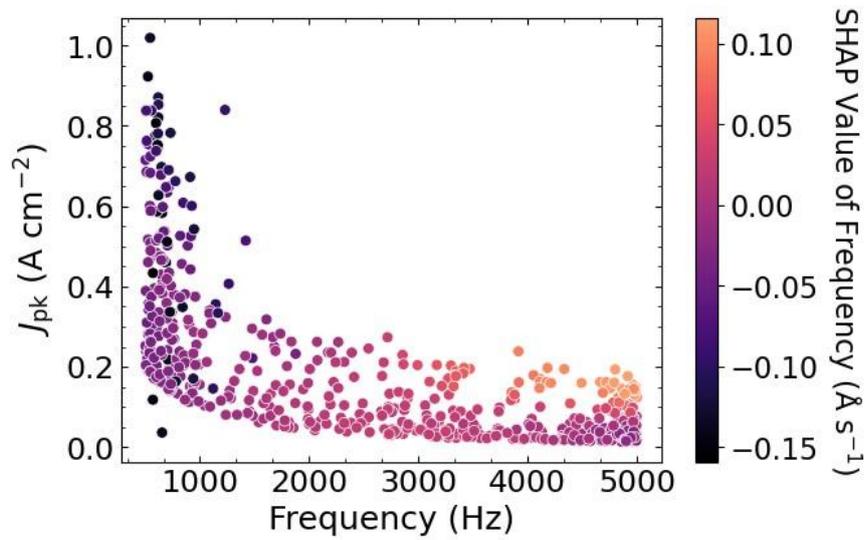

**Figure S4.** $J_{pk}$ is plotted against Frequency, highlighting their correlation; higher $J_{pk}$ values exist at lower frequency. This has to do with power control used during the sputter process, needed to maintain $J_{pk}$ at values where arcing was not observed.

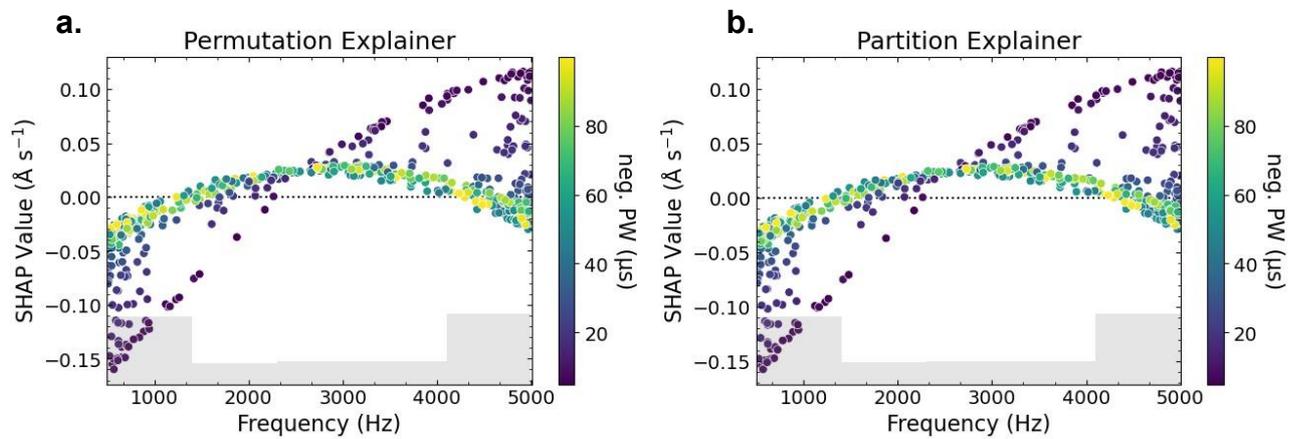

**Figure S5.** For the low pulse width, Al dataset presented in **Figure 3**, conditional SHAP explanations were evaluated to see if any interaction effects caused by correlated variables and spurious attributions could be resolved. In **a.** for a *Permutation Explainer* and **b.** a *Partition Explainer*. In both cases, the calculated SHAP values are nearly identical to the *Exact Explainer* used in the main body of this work, with only minute differences. This indicates the interaction effects observed here may have some validity, with *Frequency* representing the background plasma conditions of the process.

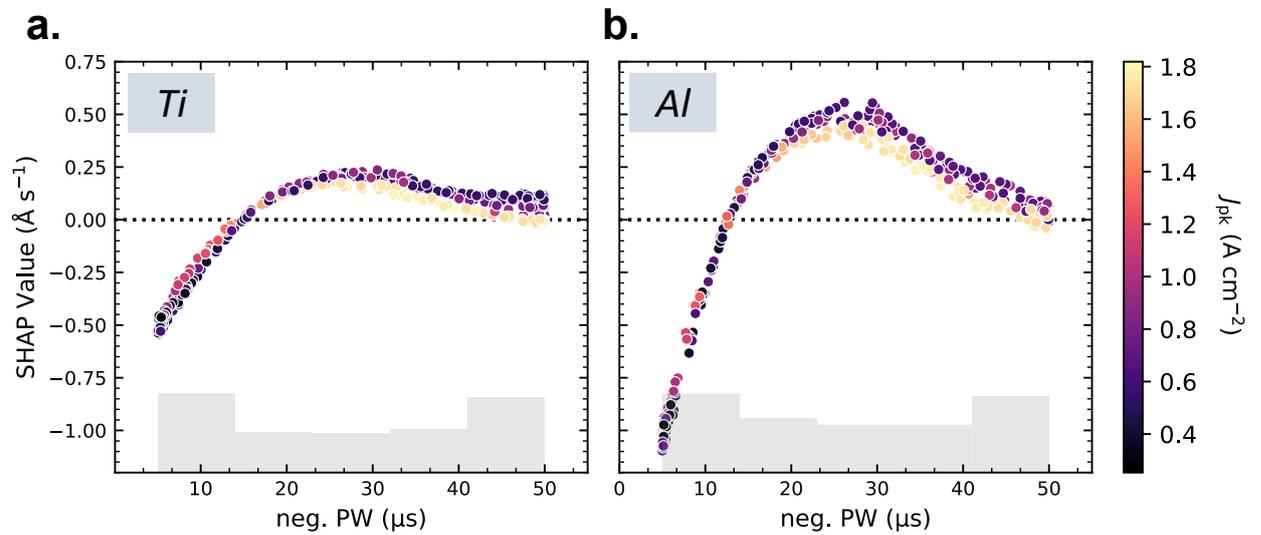

**Figure S6.** SHAP dependence plots for the low duty-cycle datasets of **a.** Ti and **b.** Al. The mode mean is indicated by a dotted line, and the distribution of data points is visible as a grey bar chart at the bottom of the graphs. A prominent peak is observed at 25 – 30 µs attributed to the reduced effect of back-attraction, as described in the main text. The color-scale indicates an interaction between the neg. PW and the $J_{pk}$. This interaction is expected as $J_{pk}$ is an indicator for ion count, for which back-attraction plays a prominent role. Counter intuitively, as $J_{pk}$ increases, the impact of the neg. PW decreases. This may be particular to the frame of the dataset, for show boundary conditions are described in greater detail in **Table S1**.

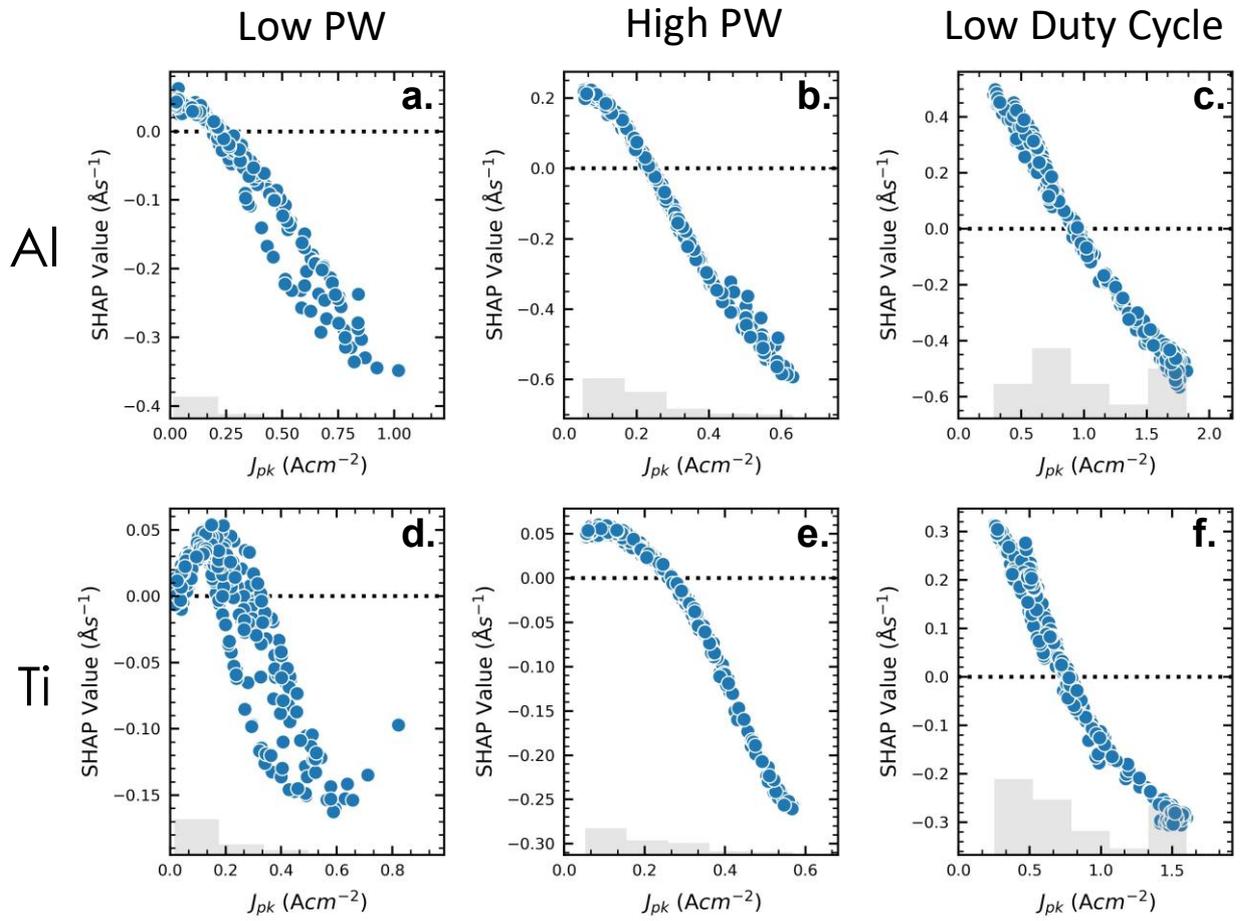

**Figure S7.** To better understand the expected relationships for individual datasets, and thus not include the potential distortion of effects caused by including heavily correlated features (e.g. Power Density), the SHAP scatter plots of $J_{pk}$ are shown above for Al (**a-c**) and Ti (**d-e**) datasets. In most cases, a relatively linear relationship appears. Of interest, a noticeable s-shape is visible for a Ti dataset collected at neg. PW (5 – 100 µs) and is expected to be an artifact from insufficient data at the extremes of the $J_{pk}$ range combined with overfitting.

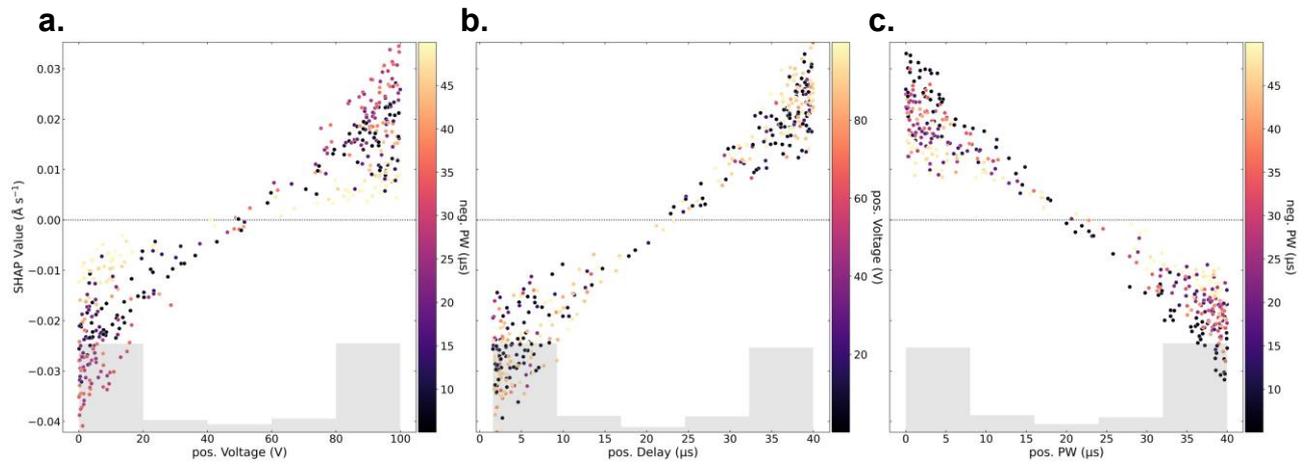

**Figure S8.** Scatter plots of the pos. pulse components for a Ti low duty-cycle dataset for **a.** pos. Voltage, **b.** pos. Delay, and **c.** pos. PW are shown above. The most prominent interaction effects (calculated with grid variance) of the pos. pulse components are plotted on the color-scale, with the quantified variance shown in **Table S1**. All parameters take on a bow-tie shape, common with parameters that are estimated to have low impact. The interactions are 'weak' and extrapolation of trends seen here cannot be done reliably. The distribution of datapoints is visualized by the grey bar chart at the bottom of each graph, with the mean of the dataset (0) indicated by a black dotted line.

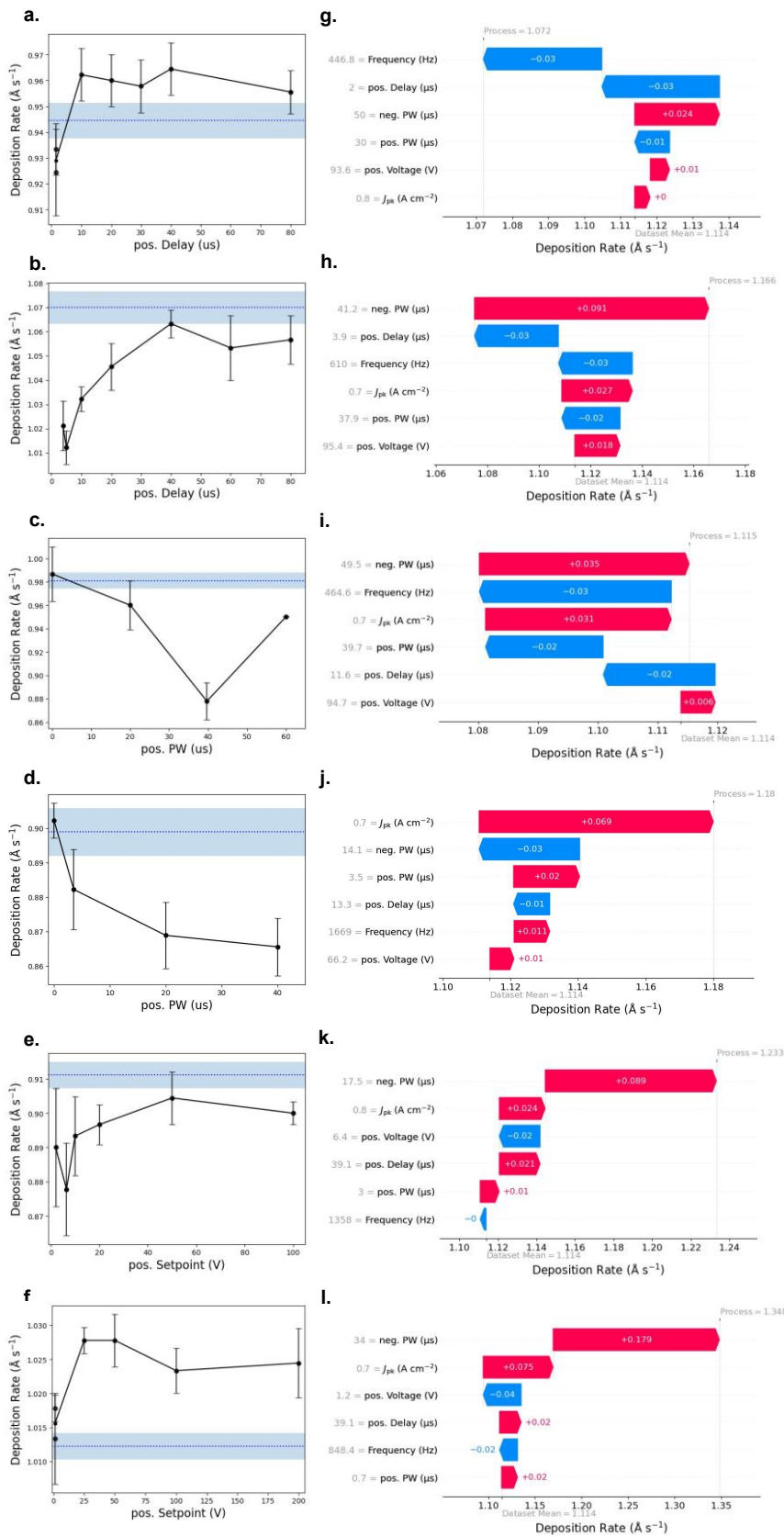

**Figure S9.** Individual process conditions were evaluated following the SHAP analysis. This was done by first evaluating where pos. pulse components were expected to have the most impact, done by comparing their SHAP values to the absolute sum of all SHAP values in the process. Two high-impact points were then chosen and their deposition rates measured while sweeping that parameter for pos. Delay (**a-b**), pos. PW (**c-d**) and pos. Voltage (**e-f**). The corresponding process conditions for each are shown to the right as waterfall plots in (**g-l**). The dotted blue line corresponds to the process condition where all pos. pulse components are turned off, to help evaluate increases in deposition rate to a purely unipolar case. In general, the trends follow those seen in the SHAP analysis, but lose their linearity, likely due to the sparse sampling conditions at the edge of the parameter space.